\documentstyle[12pt]{article}
\newcommand{\argm}{\mathop{\rm argmin}\limits}

\newcommand{\const}{\mathop{\rm const}\limits}

\begin{document}
\begin{center}

{\bf OPTIMAL ADAPTIVE NONPARAMETRIC DENOISING OF } \\

\vspace{2mm}

 {\bf MULTIDIMENSIONAL \ - \ TIME SIGNAL.}\\

\vspace{2mm}

                Eugene Ostrovsky \\
  Department of Mathematic, Soniclynx  company, \\
56209, Rosh Ha'ain, Hamelecha  street, 22, ISRAEL; \\
     E-mail: eugeny@soniclynx.net \\

\vspace{2mm}

               Leonid Sirota\\
Department of  Mathematic, Bar \ - \ Ilan University,  Ramat Gan, ISRAEL, 52900,\\
 E-mail: sirota@zahav.net.il \\

\vspace{3mm}

{\sc ABSTRACT}

\end{center}

\vspace{3mm}

 We construct an adaptive asymptotically  optimal in the classical  norm of the space $ L(2) $ of square integrable functions non \ - \ parametrical multidimensional "time"  defined signal regaining (adaptive filtration, noise canceller) on the background noise via multidimensional truncated Legendre expansion and {\it optimal experience design. } \par
 The two \ - \ dimensional case is known as a "picture processing,"  "picture   analysis" or "image processing". \par
 We offer a two version of an confidence region building,  also adaptive. \par
  Our estimates proposed by us have successfully
passed experimental tests on problem by simulate of modeled with the
use of pseudo-random numbers as well as on real data (of seismic signals etc.)
for which our estimations of the different signals were compared with
classical estimates obtained by the kernel or wavelets
estimations method. The precision of proposed here estimations is better. \par
 Our adaptive truncation may be used also for the signal and image
{\it compression}. \par

 \vspace{3mm}



{\bf 1. Statement of problem.} Let  $ V(n),  n = 16,17, \ldots  $ be a sequence
of a vector \ - \ valued  sets  (plans of experiences) in the cube
$ [-1,1]^d, \ d = 2,3, \ldots :$

$$
V(n) = \{ x_i = \vec{x}_i = \vec{x}_i(n), \}, \
\vec{x}_i \in [-1,1]^d.
$$

 At the points $ \vec{x}_i $ we observe the unknown
signal $ f = f(x), \ x \in [-1,1]^d $ on the background
noise:
$$
y(i) = f(\vec{x}_i) + \sigma \ \xi_i, \eqno(1)
$$
where the noise $ \{ \xi_i \}, \ $ errors of measurements,
is the sequence of independent (or weakly dependent) centered: $ {\bf E } \xi_i = 0 $ normed: $ {\bf Var }(\xi_i) = 1 $ random variables, $ \sigma =
\const > 0 $ is standard deviation of errors. \par
{\bf Our aim is elaboration of an adaptive asymptotically as
$ n \to \infty $ optimal in the $ L(2) $ sense signal $ f $ retaining }
$ f_n = f_n( \vec{x}): \ \Delta^2(n) \stackrel{def}{=} $
$$
{\bf E} ||f_n(\cdot) \ - \ f(\cdot)||^2 =
{\bf E} \int_{ [-1, 1]^d} |f_n(x) \ - \ f(x)|^2 \ dx \to \min_{f_n};
$$
 {\bf $ f_n = f_n(x; V(n), \{ y(i) \} ) $ is some measurement, or, in other words, estimation of a signal $ f = f(x). $ } \par
 {\bf We consider in this report only multidimensional case } $ d \ge 2. $
The one \ - \ dimensional case is consider in [3].  We notice that there are some essential differences between one \ - \ dimensional and multidimensional
cases; we will show, for example, that in the multidimensional case we need to use only optimal experience design.\par
 The multidimensional case $ d \ge 2 $ imply that our signal, more exactly,
the function on $ x, $ is not necessary to be "temporal". \par
 The adaptiveness means that our estimations do not use any apriory information
about the estimated function $ f, $ for example, information on the its class of
smoothness.\par
 On the other words, this problem is called "filtration of a signal on the
background phone", "adaptive noise canceller" or "regression problem".\par
 In the one \ - \ dimensional case $ d = 1 $ this problem was considered in many publications ([1] \ - \ [5] etc). The case $ d = 2 \ $ is known as "picture processing" or equally "image processing".  \par

\vspace{2mm}

{\bf 2. Denotations. Assumptions. Construction of our retaining.} Let
$ \vec{z} = z = \{z_j \}, j = 1,2,\ldots, d, \ z_j \in [-1,1]  $  be a $ d \ - $ dimensional vector,
$$
F(\vec{z}) = 2^{-d} \prod_{j=1}^d (1 + z_j),  \ \delta(n) =
\delta(n,V(n)) =
$$
$$
\sup_x |G_n(z) \ - \ F(z)|, \ G_n(z) = n^{-1} \sum_{i=1}^n I(x_i
< z),
$$
where
$$
I( \vec{x} < \vec{z}) = 1 \Leftrightarrow \forall j = 1,2, \ldots,d \
\Rightarrow \ x_j < z_j,
$$
and $ I(\vec{x} < \vec{z} ) = 0 $ in other case. \par
 The value, more exactly, the function $ \delta = \delta(n) = \delta(n,V(n)) $  is called {\it discrepancy} of a sequence plans $  V(n). $ \par
 We suppose that

$$
\delta(n) \le  C(1,d) [\log( n)]^d /n, \eqno(2)
$$

 Note that in the one \ - \ dimensional case the condition (2) is
satisfied even without the member $ \log^2(n) $
if $ x_i = -1 + 2 i/n $ (the uniform plan); but in general case $ d \ge 2 $ we need to use, e.g., the Niederreiter's sequences  {\it (experience design)}(see [6], p. 183 \  -  \ 202), for which the condition (2) is satisfied.\par

 It is proved also in [6], p. 251 \ -  \ 276 that for arbitrary {\it sequences }
of plans $ V = V(n) $ its discrepancy satisfies the inequality

$$
\delta(n) \ge C(2,d) \ (\log n)^{d - 1}/ n.
$$
Therefore, the Niederreiter's sequences are quasi \ - \ optimal in the sense of
minimal asymptotical as $ n \to \infty $ behavior of discrepancy $ \delta(n). $ \par
 In comparison, for the uniform
$ d \ - $  dimensional plan $ \delta(n) \asymp C(3) \ n^{-1/d}. $ \par
  It is well known that for the so \ - \ called {\it random} experience design, i.e. if the vectors $ \{ \vec{x_i} \} \ $ are random variables with the uniform distribution in the cube $ [-1, 1]^d, $

 $$
 \delta(n) \ge C(3,d) \ (\log \log n)^{1/2}/ \sqrt{n},
 $$
where $ C(3,d) $ are the random constants.\par
 Therefore, the uniform plans and the random plans are not asymptotically  optimal.\par
 Note in addition that the Niederreiter's sequences allow us to elaborate  the
 {\it sequential } estimation of signal $ f(\vec{x}). $ \par

 Further, we assume that for some $ q,Q \in (0,\infty) $

$$
\forall u \ge 0 \ \Rightarrow {\bf P}(|\xi_i| > u) \le \exp \left( - (u/Q)^q   \right). \eqno(3)
$$

 The condition (3) is satisfied, e.g., if the errors of measurements $ \{ \xi_i  \} $ have the Gaussian distribution; in this case $ q = 2. $ \par
 The consistent as $ n \to \infty $ measurement (estimation) $ Q(n), q(n) $
and  $ \gamma(n) $
of the parameters $ Q, q $ is described correspondently  in [1], [2].\par
  Further, let us denote by $ L_m(x) $ the usually {\it normed} Legendre's polynomial on the set [-1,1]. The Legendre polynomials $ P_m(x) $ are given by the well \ - \ known Rodrigues formula

$$
P_m(x)={1 \over {2^m m!} } {d^m \over {d x^m}} \bigl [(x^2-1)^m \bigr]
$$

 or, more conveniently for computation, by means of recurrent relation and initial conditions: $ P_0(x) = 1, \ P_1(x) = x, \ m \ge 1 \ \Rightarrow $
$$
(m+1) P_{m+1}(x)= (2m + 1) \ x \ P_m(x) \ - \ m \ P_{m \ - \ 1}(x)
$$

with orthogonal property:
$$
I(k,m) \stackrel{def}{=}
\int_{ \ - \ 1}^1 P_m(x)P_k(x)dx= 2/(2m+1), \ m = k,
$$
 otherwise $ I(k,m) = 0. $ We can define $ L_k(x) = P_k(x) \sqrt{k + 0.5} $
and for the multidimensional index
$ \vec{k} = k = (k(1), k(2), \ldots, k(d)), \ k(j) = 0,1, \ldots, d $
$$
\phi( \vec{k}, \vec{z}) = \prod_{j=1}^d L_{k(j)}(z(j)), \ \vec{z} =
 \{ z(j), \ j = 1,2,\ldots, d \}.
$$
 We denote  $ \nu = 2^{1/d} \ $ and for $ N \in \ (1,  \ N_d(n)) $

$$
R(N) = \{ \vec{k}: \max_j k(j) \le N \},
 \ W(N) = R( [\nu \ N]) \setminus R(N).
$$
 Hereafter $ [z] $ will denote the integer part of (positive) variable
 $ \ z.$\par
 Since the function (signal) $ f = f(x) \ $ is presumed to be
square integrable: $ f \in L(2), $ it may be expanded in the $ L(2) $ sense
as follows:

$$
f(z) = \sum_{\vec{k}} c(\vec{k}) \ \phi( \vec{k}, \vec{z}), \
\rho(N) \stackrel{def}{=} \sum_{ \vec{k} \notin R(N)}
\{ c(\vec{k}) \}^2 \to 0, \ N \to \infty.
$$

We suppose (condition $ \gamma \ ) $ that there exists a limit less
than $ 0.5: $

$$
\gamma \stackrel{def}{=}
\lim_{N \to \infty} \rho( [\nu \ N] )/\rho(N) < 1/2, \eqno(4)
$$
and will write  $ f \in K(\gamma). $ In the case when $ \gamma =
0 $ we will write $ f \in K(0). $  \par
 The condition $ \gamma $ is satisfied if, e.g., as $ N \to \infty $

$$
\rho(N) \sim C(5) N^{- \beta} S(N), \ \beta > d, \eqno(5),
$$
where $ S = S(N) $ is slowly varying as $ N \to \infty $ function;
the condition $ f \in K(0) $ is satisfied if, e.g., as $ N \to
\infty $

$$
\rho(N) \sim C(6) \ \alpha^N, \ \alpha = \const \in (0,1). \eqno(6)
$$

 The values $ \rho(N) = \rho(f,N) $ are known and well studied in the
{\it approximation theory. } Namely, $ \rho(f,N) = E^2_{N}(f), $ where
$ E_{N}(f) $ is the error of the best approximation of $ f $
by the algebraic polynomials of each power not exceeding $ N $ in the $ L(2) $
distance and are closely connected with module of continuity of the form

$$
\omega_{\psi}^m (f, t) = \sup _{|h| \le t} || \Delta^m_{h, \psi} f||, \
 \Delta^m_{h, \psi} f(x) \stackrel{def}{=}
$$
$$
\sum_{l=0}^m ( \ - \ 1)^l m! \ f(x + (0.5 m \ - \ l) \ h \ \psi(x)) /(l! \ (m \ - \ l)!),
$$

$$
\psi(x) = (1 \ - \ x^2)^{0.5}, \ f(x+y) = f( \min(x + y,1) ), y > 0; \ f(x + y) =
$$
$ f( \max(x + y), -1) \ $ if $ y < 0; \ m = 0,1,2,\ldots; \  h = \vec{h} =
(h(1), h(2), \ldots, h(d)); \ |h| = \max_j |h(j)|. $ \par

 For instance, see ([7]), $ \rho(f,N) \asymp N^{-2m} $ if and only if

$$
\omega_{\psi}^m (f, t) \asymp  \ t^m |\log(t)|^{0.5}, \ t \in (0, 0.5].
$$

 {\bf Remark.} The condition $ \beta > d $ or more general assumption
$ \gamma  < 0.5 $  is necessary still in the case $ d = 1 \ $ ([1], [2]). \par

 We can estimate the coefficients $ c(k) $ as follows:

$$
c(n, \vec{k})  = n^{-1} \sum_{i=1}^n y(i)
\ \phi( \vec{k}, \vec{x}_i).
$$

Let us define $  N_d(n) = \left[ n^{1/(d+1)} \ (\log n)^{-2d/(d+1) } \right], \
\tau(N) = \tau(N,n) = $

$$
 \sum_{ \vec{k} \in W(N)} [c(n, \vec{k})]^2, \  N(n)  =
 \argm \{ \tau(N,n), N \le N_d(n) \},
$$

$$
f_n = f_n( \vec{x}) = \sum_{ \vec{k} \in R(N(n))} c(n, \vec{k}) \
\phi( \vec{k}, \vec{x}). \eqno(7)
$$
 The function $ f_n = f_n( \vec{x}) $ {\it represented our adaptive measurement
of an unknown signal } $ f = f(\vec{x}). $  It may be proved that our signal measurement
 $ f_n $ is optimal {\it in order } as $ n >> 1 $ in $ L(2) $ norm under conditions (3) and (5) in the minimax sense.  \par

\vspace{2mm}

{\bf 3. Properties of our estimation. Main result.}  We can obtain
after hard calculations alike to [2], [3] that as $ n \to \infty $

$$
{\bf E} ||f_n \ - \ f||^2
\sim \min_N \left( \rho(N) \cdot [ 1 \ - \ \gamma] +   \sigma^2 N^d/n  \right);
$$
therefore in the case if $ \gamma = 0 $ our estimation
$ f_n(\cdot) $ is {\it asymptotical optimal } in the $ L(2) $ sense.\par

 In the case if $ \gamma \in (0, 0.5) $ we can modify our estimation (7) in order to obtain optimal measurement of
$ f(\cdot) \ $ as follows. Instead the functional $ \tau $ we
introduce its so \ - \ called {\it penalty } modification:

$$
\theta(N) = \tau(N) \ - \ \gamma(n) \ \sigma^2(n) \ N/n \eqno(8)
$$
and define as a modified, asymptotically optimal in $ L(2) $ sense estimation for the function $ f $ the function $ g_n = $

$$
g_n(\vec{x}) = g_n( \vec{x}; V(n), \{ y_i \} ) =
 \sum_{ \vec{k} \in R(M(n))} c(n, \vec{k}) \
\phi( \vec{k}, \vec{x}), \eqno(9)
$$

$$
 M(n) =  \argm \{ \theta(N,n), \ N \le N_d(n) \}.
$$

 Here $ \gamma(n), \ \sigma^2(n), \ q(n), \ Q(n)  $ etc. are correspondently  consistent estimations of parameters $ \gamma, \ \sigma^2, \ q, \ Q $
estimation, described in [1], [2]. \par
 For instance,

$$
\sigma^2(n) = \sum_{i=1}^n \left[ f_n \left(\vec{x}_i \right) \ - \ y(i) \right]^2 /(n \ - \ N^d + 1). \eqno(10)
$$

{\bf 4. Confidence region (c.r.). } We want build in this section the c.r. for
$ f(\cdot) $ in the $ L(2) $ sense. As a first approximation we can offer the following approach. With probability tending to one as $ n \to \infty $ the following inequality holds:

$$
||f_n \ - \ f ||^2 \le Q^2(n) \ \tau(N(n))/(1 - \gamma(n)).
$$

 For the more exact c.r. building we proved that $ ||f_n \ - \ f ||^2 \le $

$$
Q^2(n) \ \tau(N(n))/(1 - \gamma(n)) \times
\left[1 + C(\gamma) \ \zeta \  (\log \log n)^{2/r} / n \right],
$$

$ r = 2 d q(n)/(q(n) + 4d), q(n) \in (0,2); \
r =  d q(n)/(q(n) + d), q(n) > 2, $

where the non \ - \ negative random variable $ \zeta $ is such that for all
positive values $ u  > 0 $

$$
\sup_n {\bf P} \left( \zeta  > u \right) \le \exp \left( - u^{r/2} \right)
\eqno(11)
$$
("exponential confidence region"). \par
{\bf 5. Optimal adaptive denoising in other norms.}
 We can consider instead the $ L(2) $ norm also some more strong norms, (in
order to improve the sensitivity of our method,)   for
example,  $ L(p) $ norm or the uniform norm $ L(\infty) $ in the space of continuous functions $ C[\ - \ 1,1] $ etc.:

$$
\Delta_p(h_n, f) = || h_n \ - \ f||_p \stackrel{def}{=}
{\bf E} \left[ \int_{[ \ - \ 1,1]^d } |h_n(x) \ - \ f(x)|^p \ dx  \right]^{1/p};
$$

$$
\Delta_{\infty}(h_n, f) = {\bf E} \sup_{x \in [ \ - \ 1,1]^d}|h_n(x) \ - \ f(x)| = \lim_{p \to \infty} \Delta_p(h_n, f),
$$
where $ h_n(\cdot) $ is some estimation (measurement) of signal
$ f(\cdot). $ \par
 But for consistent and optimal measurement in these spaces we need to use the
so \ - \ called Vallee \ - \ Poissin improvement of $ g_n(\cdot). $ Namely,
let us denote

$$
|\vec{k}| = |k| = \max_{j = 1,2,\ldots,d} |k(j)|, \  m_{\infty} =
m_{\infty}(n) = N[n/\log(n]
$$
in the case $ p = \infty $ and
$$
 m_{p}  = m_{p}(n) = N(n)
$$
in the case $ p < \infty. $ \par

 We define the {\it Vallee \ - \ Poissin modified} coefficients
 $$
 d(\vec{k}, n) = d(k,n) = d_p(k,n) = c(k,n), |k| < m_p;
 $$

$$
d_p(k,n) = c(k,n)(\nu N(n) \ - \ |k|)/(\nu N(n) \ - m_p(n)), |k|
\in [m_p(n), \nu N].
$$
 As the estimation $ h_n(\cdot) = h^{(p)}_n(\cdot) = h^{(p)}(\cdot) $ of a
signal $ f(\cdot) $ we offer the following improvement of the estimation $ g_n:$

$$
h^{(p)}_n( \vec{x} ) = \sum_{ \vec{k} \in R(\nu \ N(n))} d_p(n, \vec{k}) \
\phi( \vec{k}, \vec{x}).
$$

 This estimation $ h_n(\cdot) = h^{(p)}_n(\cdot) $ of a signal $ f $ is optimal in
order as $ n \to \infty $
in each space $ L(p), \ p \in (2,\infty] $ norms. \par

  For the simple building of confidence region in the $ L(p)$ norms we proved also  that  as with probability tending to one as $ n \to \infty $
$$
  ||h^{(p)}_n \ - \ f||_p \le C_7(p, q(n), \gamma(n)) \ Q(n) \
\tau(m_p(n))/(1 \ - \ \gamma(n)), \ p < \infty,
$$

and
$$
 ||h^{( \infty) }_n \ - \ f||_{\infty} \le C_8(q(n), \gamma(n))
 \ \tau(m_{\infty}(n)) \ Q(n)/ (1 \ - \ \gamma (n) ).
$$

{\bf 6. Proofs. } Notice that the complete mathematical proof of our
assertions used the modern {\it martingale theory,} for instance the exponential bounds for tails
 of distribution in the Law of Iterated Logarithm (LIL) for martingales, as in the one \ - \ dimensional case considered in [3]; {\it theory approximation} [12] and {\it theory of Banach spaces of random variables } [13] etc. \par
 Our proof is alike to the proofs in one \ - \ dimensional case [3]; we must
explain only briefly some new essential moments. \par

{\bf A. } Let us denote

$$
A(n,N) = \rho(N) + \sigma^2 N^d/n, \ A(n) = \min_{N = 1,2,\ldots} A(n,N);
$$

$$
B(n,N)= \sum_{k \in W(N)} (c(k))^2 + \sigma^2 N^d/n \sim \rho(N) \
(1 \ - \ \gamma) + \sigma^2 N^d/n;
$$

$$
B(n) = \min_N B(n,N) = \min_{N \le N_d(n)} B(n,N);
$$

$$
N^0 = N^0(n) = \argm_{N = 1,2,\ldots} B(n,N) =
\argm_{N \le N_d(n) } B(n,N).
$$
 It follows from the condition $ (\gamma) $ that as $ n \to \infty $
$$
A(n,N) \asymp B(n,N), \ A(n) \asymp B(n),
$$
and, by virtue of condition $( \gamma ) $
$$
N^0 \asymp \argm_{N = 1,2,\ldots} A(n,N) \asymp \argm_{N \le N_d(n)} A(n,N).
$$
 The value $ (A(n))^{1/2} $ is asymptotical optimal in $ L(2) $ sense as
$ n \to \infty $ speed of convergence of an arbitrary, i.e. not necessary
to be adaptive, estimations of the function $ f(\cdot) \ $ [14].

{\bf B.} We can write further:

$$
c(k,n)  \sim c(k) + n^{-1/2} \theta_k(n) + \eta(k,n),
$$

where  the deterministic variables

$$
\eta(k,n) = n^{-1} \sum_{i=1}^n f( \vec{x_i}) \ \phi(\vec{x_i}) \ - \ c(k)
$$
are errors of Fourier \ - \ Legendre coefficients $ \{ c( \vec{k}) \} $
numerical computing by means of plan (set) $ V(n) $ with equal weights. \par

 We obtain after the $ d \ - $ times integration  "by parts" using the known   properties of Legendre's polynomials and the condition $ (\gamma): $

$$
\eta(k,n) = \int_{ [ \ - \ 1, 1]^d}f(x) \ \phi(k,x)
 \ d \left( G_n(x) \ - \ F(x) \right);
$$

$$
|\eta(k,n)| \le C(\gamma,d) \ \sup_x |G_n(x) \ - \ F(x)| \
\left(|k|^d + 1) \right) =
$$

$$
C ( \gamma,d) \ \delta(n, V(n)) \ ( |k|^d + 1 );
$$

$$
\Sigma^2_{\eta} \stackrel{def}{=} \sum_{k \in R( \nu \ N ) } |\eta(k,n)|^2 \le
C(\gamma,d) \ \log^{2d}(n) \ N^{2 d + 1} \ n^{-2}.
$$

 Since the value $ N(n) $ belong to the segment $ (1, N_d(n)), $
we conclude  after simple computations
that the sum $ \Sigma^2_{\eta} $ not exceeded the value $ C \ N^d/n
\le B(n,N) \sim \tau(n,N).$ \par

{\bf C.} We have:
$$
\theta_k(n) = \sigma \ n^{-1/2} \sum_{i=1}^n \xi_i \phi_k(x_i).
$$

 It follows from the multidimensional CLT that the variables
$ \ \{ \theta_k(n) \} $ for all the values $ k = \vec{k} $ as $ \ n \to \infty $
are asymptotically Gaussian distributed and independent:

$$
{\bf Var} [\theta_k(n)] = n^{-1} \sum_{i=1}^n \sigma^2 L^2_k(x_i)
\to \sigma^2 \ \int_{[ \ - \ 1, 1]^d } \phi^2_k(x) \ dx = \sigma^2;
$$

$$
{\bf E} \theta_k(n) \theta_l(n)  = \sigma^2 n^{-1} \sum_{i=1}^n \phi_k(x_i)
\phi_l(x_i) \to \sigma^2 \ \int_{ [ \ - \ 1, 1]^d } \phi_k(x) \phi_l(x) \ dx = 0, \ k \ne l.
$$
 Following, the variables $  \{ \theta_k(n) \} $
are asymptotically independent and have approximately the normal distribution:
$$
Law( \ c(k,n) \ ) \asymp N(c(k),\sigma^2/n),
$$
or equally
$$
c(k,n) = c(k) + \sigma \epsilon_k/\sqrt{n}, \ Law(\epsilon_k) \asymp N(0,1)
$$
and also $ \{\epsilon_k\} $ are asymptotically independent. Therefore,
$ \tau(n,N) \asymp $

$$
\sum_{k \in W(N) } |c(k)|^2 + 2 \ n^{-1/2} \ \sigma
\sum_{ k \in W(N) } c_k \epsilon_k + \sigma^2 \ n^{-1}
\sum_{k \in W(N)} \epsilon_k^2 =
$$

$$
\sum_{k \in W(N) } |c(k)|^2 + 2 \ n^{-1/2} \ \sigma
\sum_{ k \in W(N) } c_k \epsilon_k + \sigma^2 n^{-1} N^d  \ + \
 \sigma^2 \ n^{-1} \sum_{k \in W(N)} (\epsilon_k^2 -1);
$$

$$ 	
{\bf E} \tau(n,N) \asymp B(n,N), \ \ {\bf Var} [\tau(n,N) ]
\asymp B(n,N)/n,
$$
and hence

$$
N \to \infty, N/n \to 0 \ \Rightarrow \sqrt{ {\bf Var} [\tau(n,N)]}/
{\bf E}\tau(n,N) \to 0.
$$
 Note that the conditions  $ \gamma < 1/2  $ and (2) was used and is essential which is common in statistical research. \par

{\bf D. } It follows from our considerations that there are some grounds to conclude
$$
\tau(n,N) \stackrel{a.s}{\asymp} {\bf E} \tau(n,N) \asymp A(n,N),
$$
thus,
$$
N(n) = \argm_{N \le [n^{1/d}/3]} \tau(n,N) \sim \argm_{N \le [n^{1/d}/3]}
{\bf E} \tau(n,N) = N^0(n).
$$
 Since the our adaptive value (random!) of amount summands $ N(n) $ is near
to the optimal  value $ N^0(n), $ (not adaptive,) our estimation (measurement)
 is also optimal.\par
  More exactly,  we can write as a first approximation without the members
$ \{ \eta(k,n) \} $ calculations: $ c(k,n) = c_k + $
$$
 \sigma \ n^{-1} \sum_{i=1}^n \xi_i \phi(k,x_i);  \ (c(k,n))^2 =  c^2(k) +
$$

  $$
   \sigma^2 \ n^{-2} \ \sum_{i=1}^n  \phi^2(k,x_i) +
   2 \ \sigma \ n^{-1} \sum_{i=1}^n  c(k) \ \xi_i \ \phi(k,x_i) +
  $$

   $$
   \sigma^2 \ n^{-2} \sum_{i=1}^n \left( \xi^2_i \ - \ 1 \right)
  \phi^2(k,x_i) +
    $$

    $$
  2 \ \sigma^2 \ n^{-2} \sum \sum_{1 \le i < j \le n} \xi_i \ \xi_j
 \  \phi(k,x_i) \ \phi(k,x_j).
    $$

  We have for the variables $ \tau(n,N) $ (and further for the variables
 $ \Delta^2 = \Delta^2(n,N) = ||\hat{f} - f||^2 \ ): $  $ \tau(n,N) = $

$$
\left[ \sum_{k \in W(N)} c^2_k + \sigma^2 n^{-1} \sum_{k \in W(N)}
n^{-1} \sum_{i=1}^n \phi^2(k,x_i) \right] +
$$

 $$
  2 \ \sigma \ n^{-1} \sum_{i=1}^n \xi_i \sum_{k \in W(N) } c(k) \
\phi(k,x_i) + \tau_2,
 $$

 $$
\tau_2 = \sigma^2 \ \left[ n^{-1} \sum_{i=1}^n \left(\xi^2_i \ - \ 1 \right)
\sum_{k \in W(N) } \phi^2(k,x_i) \right] +
 $$

 $$
\sigma^2 \ \left[ 2 \ n^{-1} \sum \sum_{1 \le i < j \le n} \ \xi_i \ \xi_j \
 \sum_{k \in W(N)} \phi(k,x_i) \ \phi(k,x_j) \right].
 $$

  Note that the sequences  of a view
$ \eta_1(n) = \sum_{i=1}^n b(i) \ \xi(i), $

$$
  \eta_2(n) = \sum_{i=1}^n  b(i) \ (\xi^2_i \ - \ 1)
$$
and
$$
\eta_3(n) = \sum \sum_{1 \le i < j \le n} b(i,j) \ \xi_i \ \xi_j,
$$
where $ \{ b(i) \}, \{b(i,j) \} $ are a non-random sequences, with the second
component $ F(n) = \sigma \left( \{ \xi_i \}, i = 1,2,\ldots,n \right), $ i.e.
$ \{ \eta_s(n), F(n) \}, s = 1,2,3;  \ \{ F(n), \ n = 1,2,3, \ldots \} $ is
the natural sequence (flow) of sigma-algebras (filtration),
are martingales. \par
  Using the main result of paper [15], devoted to the Law of Iterated Logarithm
for martingales, and repeating the considerations of the article [3] about the one \ - \ dimensional case, we obtain desired.\par

{\bf 7. An example.} Suppose for some constants
$ \beta > 0.5, \ K \in (0, \infty) $ as $ N \to \infty $
$$
\rho(N) \sim K^{d + 2 \beta} \ N^{-2 \beta } /(2 \beta).
$$

Then we have for the estimation $ g_n(\cdot) $ as $ n \to \infty:$
$ {\bf E} ||g_n  \ - \  f ||^2 \sim  $

$$
K^d \ n^{-2 \beta/(2 \beta + d)} \ \sigma^{4 \beta/(d + 2 \beta) } \times
d^{2 \beta/(2 \beta + d)} \ \left[ \frac{1}{2 \beta} + \frac{1}{d} \right].  \eqno(12)
$$

 Thus, the rate of convergence $ g_n \to f $ in the $ L(2) $ sense
is optimal ([2]).\par
 Note that by construction of our estimations we do not use the (unknown, as usually) parameters $ K, \beta \ $ (adaptiveness).\par
  Notice in conclusion that the estimates proposed by us have successfully
passed experimental tests on problem by simulate of modeled with the
use of pseudo-random numbers as well as on real data (of seismic signals etc.)
for which our estimations of the different signals $ f $ were compared with
classical estimates obtained by the kernel or wavelets
estimations method. The precision of proposed here estimations is better. \par
{\bf 8. The computation complexities. }  The amount  $ AM(n) $ of an elementary  operation and square roots calculations of offered algorithm,
{\it if we will use  the so \ - \ called Fast Legendre Transform (FLT) [8]}
is equal to
$$
AM(n) \asymp \left(C(d) \ n \ \log_2 n \right)^d.
$$
 Recall (see [9]) that the amount of these operations by using the
classical Fast Fourier Transform (FFT), even in the $ d \ - $ dimensional
 case is equal to $ C(d) \ n \ \log_2 n. $ \par
 The advantage of our estimations in comparison to the trigonometric
estimations [2] is especially in the case when the estimating function
$ f(\cdot) $ is not periodical: $ f(-1,-1,\ldots,-1) \ne f(1,1,\ldots,1). $ \par
{\bf 9. Detection of signal.} We can to use our adaptive c.r. for construction
a test for {\it presence (detection)} of a signal. Namely, let us consider the following statement of hypothesis verification problem: $ H_0 = \{ f = 0 \} $ (the absence of signal) versus alternative $ H_1 = \{ f \ne 0 \} $ (the presence of the signal). \par
 As long as the hypothesis  $ H_0 $ may be reformulated as $ H_0 =
\{ ||f||^2 = 0 \} $ and the counterhypothesis has a view
$ H_1 = \{ ||f||^2 > 0 \}, $ we can offer the following test. \par
 Let $ \delta, \ \delta \in (0, 1/3) $ be some "small" number, for example, 0.05
 or 0.01 etc., such that the value $ \delta $ is allowed level of a first kind:

$$
{\bf P}(H_1/H_0) \le \delta. \eqno(13)
$$

{\it Our test $ \phi $ may be defined as follows: } $ \phi = 1 $ if and only if

$$
 ||f_n||^2 \ge K(\delta),
$$
and $ \phi = 0 $ in other case.\par
 Here $ \phi(\cdot) $ denotes the number of our solution: we conclude $ H_1 $
in the case if $ \phi = 1 $ and $ H_0 $ in other case. \par
  Here the value $ K(\delta) = K_n(\delta) $ may be {\it computed} from (11) and (13),
 on the basis of equality
 $$
 {\bf P_0} \left( ||f_n||^2 > K(\delta) \right) \approx \delta.
 $$
  The notation $ {\bf P_0 }(A), \ A $ is an arbitrary event, denotes as usually
 the probability of $ A $ calculated under assumption of {\it absence }
 of  signal $ f. $ \par
  In detail
 $$
 {\bf P} \left( \frac{Q^2(n) \ \tau(N(n))}{ 1 \ - \ \gamma(n) }
 \left[ 1 + C(\gamma) \ \zeta \ \frac{ (\log \log n)^{2/r} }{n} \right] > K(\delta)  \right) \approx \delta.
 $$

 We find, solving the last equality relative $ K_n(\delta); $

 $$
 K_n(\delta) \approx \frac{Q^2(n) \ \tau(N(n)) }{ 1 \ - \ \gamma(n) }
 \left( 1 + C(\gamma) \ \frac{| \log \delta |^{2/r} \ (\log \log n)^{2/r} }{n} \right). \eqno(14)
 $$

 The advantage of offered here test versus, e.g., the tests  described
in [10], [11] etc. is following. Our procedure is non \ - \ parametrical and
adaptive, but is still consistent and asymptotically optimal in the $ L(2) $ sense.\par

 \vspace{2mm}

\begin{center}

            {\bf  References}\\

 \end{center}

1. Golubev G., Nussbaum M.  {\it Adaptive spline Estimations in the nonparametric regression Model.}  Theory Probab. Appl., 1992, v. 37 $ N^o $ 4, 521 - 529.\\
2. Ostrovsky E., Sirota L. {\it Universal adaptive estimations and confidence  intervals in the non-parametrical statistics.} Electronic Publications, arXiv.mathPR/0406535 v1 25 Jun 2004.\\
3. Ostrovsky E., Zelikov Yu. {\it Adaptive Optimal Nonparametric Regression and Density Estimation based on Fourier \ - \ Legendre
Expansion. } Electronic Publication, arXiv:0706.0881v1 [math.ST]
6 Jun 2007. \\
4. Donoho D.  {\it Wedgelets: nearly minimax estimation of edges.}  Annals of  of Statist., 1999, v. 27 b. 3 pp. 859 - 897.\\
5. Donoho D.  {\it Unconditional bases are optimal bases for data compression and for statistical estimation.} Applied Comput. Harmon. Anal., 1996, v. 3 pp. 100 - 115.\\
6.Keipers R., Niederreiter W. {\it The uniform Distribution of the
Sequences. } Kluvner Verlag, Dorderecht, 1983.\\
7. Dai F., Ditzian Z., Tikhonov S. {\it Sharp Jackson inequalities.} Journal
of Approximation Theory. 2007, doi:10.1016/j.jat.2007, 04.015.\\
8. Noullez A., Vergassola M. {\it A fast Legendre transform algorithm and
Applications to the adhesion model.} Journal of Scientific Computing, Springer,
1994, v. 9, $ N^o 3, $ p. 259 \ - \ 281.\\
9. Frigo M. and Jonson S.G. {\it The Design and Implementation of FFTW3.}
2005, Proceedings of the IEEE, {\bf 93, } p. 216 \ - 231.\\
10. Rolke W.A., Lopez A.M. {\it A Test for the Presence of a Signal.} Electronic
Publication, arXiv:0807.2149v.1 [physics. data \ - \ an] 14 Jul 2008. \\
11. Rolke W.A., Lopez A.M., Conrad J. {\it Limits and Confidence Intervals
 in the Presence of Nuisance Parameters}, Nuclear Instruments and Methods, A.,
551/2 \ - \ 3, 2005, pp. 493 \ - 503, physics/0403059. \\
12. De Vore R.A., Lorentz G.G. {\it Constructive Approximation.} Springer Verlag,  1993. \\
13. Ostrovsky E. {\it Exponential estimations for random fields.} Publishing
House OINPE, Moscow \ - \ Obninsk, 1999 (in Russian). \\
14. Ibragimov I.A., Khasminsky R.Z. {\it On the quality boundaries of nonparametric estimation of regression.} Theory Probab. Appl., 1982, v. 21 b. 1, 81 - 94.\\
15. Ostrovsky E., Sirota L. {\it Exponential Bounds in the Law of Iterated Logarithm for Martingales. } Electronic publications, arXiv:0801.2125v1 [math.PR] 14 Jan 2008.\\

\end{document}